\documentclass[onecolumn,amsmath,amssymb,12pt,superscriptaddress,nofootinbib]{revtex4}
\pdfoutput=1

\usepackage[latin1]{inputenc}
\usepackage[english]{babel}
\usepackage{amssymb}
\usepackage{amsmath}
\usepackage{amsthm}
\usepackage[]{graphicx}
\usepackage[]{subfigure}
\usepackage{tensor}
\usepackage{color}
\usepackage{cancel}
\usepackage{setspace}
\usepackage{fancyhdr}
\usepackage{framed}
\usepackage{tikz}
\usepackage[bookmarks,linktocpage, colorlinks=true, plainpages = false, citecolor = blue,  linkcolor=blue, urlcolor = blue, filecolor = blue]{hyperref}

\begin{document}

\allowdisplaybreaks
\begin{titlepage}

\title{Inconsistencies of the New No-Boundary Proposal 
%\vspace{.3in} \\ Comment on ``Damped Perturbations in the No-Boundary State'' by Diaz Dorronsoro  {\it et al.}  (arXiv:1804.01102) 
\vspace{.3in}}

\author{Job Feldbrugge}
\email{jfeldbrugge@perimeterinstitute.ca}
\affiliation{Perimeter Institute, 31 Caroline St N, Ontario, Canada}
\author{Jean-Luc Lehners}
\email{jlehners@aei.mpg.de}
\affiliation{Max--Planck--Institute for Gravitational Physics (Albert--Einstein--Institute), 14476 Potsdam, Germany}
\author{Neil Turok}
\email{nturok@perimeterinstitute.ca}
\affiliation{Perimeter Institute, 31 Caroline St N, Ontario, Canada}

\begin{abstract}
\vspace{.3in} \noindent In previous works, we have demonstrated that the path integral for {\it real, Lorentzian} four-geometries in Einstein gravity yields sensible results in well-understood physical situations, but leads to uncontrolled fluctuations when the ``no boundary" condition proposed by Hartle and Hawking is imposed. In order to circumvent our result, new definitions for the gravitational path integral have been sought, involving specific choices for a class of {\it complex} four-geometries to be included. In their latest proposal, Diaz Dorronsoro {\it et al.}~\cite{DiazDorronsoro:2018wro} advocate integrating the lapse over a complex circular contour enclosing the origin. In this note we show that, like their earlier proposal, this leads to mathematical and physical inconsistencies and thus cannot be regarded as a basis for quantum cosmology. We also comment on Vilenkin and Yamada's recent modification of the ``tunneling" proposal, made in order to avoid the same problems.  We show that it leads to the breakdown of perturbation theory in a strong coupling regime. 
\end{abstract}
\maketitle

\end{titlepage}

\tableofcontents

%%%%%%%%%%%%%%%%%%%%%%%%%%%%%%%%%%%%%%%%%%%%%%%%%%%%%%%%%%%%%%%%%%%%%%%%

\section{Introduction}

The no-boundary proposal of Hartle and Hawking \cite{Hartle:1983ai} has been an influential idea in theoretical cosmology for more than three decades, and with good reason: it puts forth a proposal for the initial state of the universe, from which -- assuming some set of physical laws -- everything else is supposed to follow. If true, it would do no less than explain the origin of space and time. What is more, the proposal involves only semi-classical gravity, {\it i.e.}, a theoretical framework already within reach of contemporary physics, without requiring the development of a full theory of quantum gravity. Given the promise and magnitude of this claim, it should be analyzed with great care. In previous works \cite{Feldbrugge:2017kzv,Feldbrugge:2017fcc} we attempted to put the no-boundary proposal on a sound mathematical footing by defining the gravitational path integral more carefully. Unfortunately,  we found as a consequence that the no-boundary proposal leads to a universe with fluctuations which are out of control. Our work led Diaz Dorronsoro  {\it et al.}\ to propose a new definition of the no-boundary proposal, involving an inherently complex contour in the space of four-metrics, {\it i.e.}, one which cannot be deformed to an integral over real four-metrics and hence has no geometrical interpretation. In particular, they chose to integrate the lapse $N$ over a complex contour running below the origin in the complex $N$-plane, from negative to positive infinite real values~\cite{DiazDorronsoro:2017hti}. In follow-up work~\cite{Feldbrugge:2017mbc}, we demonstrated the inconsistency of this proposal. Very recently, Diaz Dorronsoro {\it et al.}\ have proposed yet another definition of the no-boundary proposal, this time in a particular truncation of Einstein gravity and taking instead a complex contour for the lapse which encircles the origin~\cite{DiazDorronsoro:2018wro}. In this note we show that this latest incarnation of the no-boundary idea also leads to physical and mathematical inconsistencies. 

%%JLL_NEW
The instability that we demonstrated applies equally well to the tunneling proposal developed by Vilenkin starting around the same time as the no-boundary proposal~\cite{Vilenkin:1982de,Vilenkin:1984wp,Vilenkin:1994rn}. After we posted the original preprint version of this paper, Vilenkin and Yamada proposed a modification of the tunneling proposal in an attempt to rescue it~\cite{Vilenkin:2018dch}. Their new tunneling proposal involves the addition of a boundary term to the action, which has the consequence of selecting a different perturbation mode. As we explain in appendix \ref{AppB}, this amendment unfortunately introduces a strong coupling problem and, at present, its purported consequences cannot therefore be trusted. 

\section{Physical Motivation}

The path integral over four-geometries provides a well-motivated framework for the study of semi-classical quantum gravity. In analogy with Feynman's path integral formulation of quantum mechanics, one attempts to define transition amplitudes between two three-geometries $h_{ij}^{(0)}, h_{ij}^{(1)}$ by summing over all four-geometries that interpolate between the initial $h_{ij}^{(0)}$ and final boundary $h_{ij}^{(1)}$, {\it i.e.},
\begin{align} \label{pathintegral}
G[h_{ij}^{(1)};h_{ij}^{(0)}] = \int_{\partial g = h_{ij}^{(0)}}^{\partial g = h_{ij}^{(1)}} {\cal D} g e^{i S[g] /\hbar}\,,
\end{align}
where $g$ denotes the four-metric. In this note,  as in the work of Diaz Dorronsoro  {\it et al.}, we study a simplified model in which $S[g]$ is taken to be the usual action for Einstein's theory of gravity plus a positive cosmological constant $\Lambda.$

In our previous works~\cite{Feldbrugge:2017kzv,Feldbrugge:2017mbc} we demonstrated that, somewhat to our surprise, the path integral, over {\it real, Lorentzian} four-geometries yields well-defined and unique results {\it as it stands}, when evaluated semiclassically and in cosmological perturbation theory, {\it i.e.}, when we treat the four-geometry as a homogeneous, isotropic background with small, but otherwise generic, perturbations. In contrast, we found the path integral over Euclidean four-geometries (as originally advocated by Hartle and Hawking \cite{Hartle:1983ai}), even at the level of the homogeneous, isotropic background, to be a meaningless divergent integral. The key to our work was the use of Picard-Lefschetz theory, a powerful mathematical framework that allows one to rewrite highly oscillatory and only conditionally convergent integrals (such as \eqref{pathintegral} turns out to be) as absolutely convergent integrals. To do so, one regards the integral (\ref{pathintegral}) as being taken over the subspace of real, Lorentzian metrics in the space of complex four-metrics.  Cauchy's theorem, and Picard-Lefschetz theory, then allows one to deform the original, real integration domain into a complex domain consisting of one or more steepest descent thimbles. Each of these yields an absolutely convergent integral: their sum equals the original integral. Note that the analytical continuation to complex metrics and the deformation to steepest descent thimbles, are merely a convenient calculational tool, used to evaluate the original, uniquely defined but only conditionally convergent integral using steepest descent methods. To make this point very clear, we provide in appendix \ref{AppA} an explicit proof of the convergence of integrals of the type we encounter, taken over real values of the lapse function $N$. Thus, one can prove the original integrals exist, and only subsequently use Picard-Lefschetz theory to evaluate them.

One frequently raised question is the range over which the lapse $N$ should be integrated over in the path integral. The Lorentzian four-geometries we consider may be parameterized with the line element $-N^2(t,x) dt^2 +h_{ij} (t,x)dx^i dx^j$, where $0\leq t\leq 1$ is a good time-like coordinate, {\it i.e.}, a one to one, invertible map from the manifold into the closed unit interval. The lapse $N$ accounts, for example, for the proper time interval $\tau$ between two spacetime points $(t_1, x^i$) and $(t_2,x^i)$, both at fixed $x^i$:  one has $\tau=\int_{t_1}^{t_2} N(t,x^i) dt$. Note that the coordinate $t$ already defines an orientation for the integral: the lapse $N$ is simply a local rescaling, which must therefore be taken strictly positive as long as the coordinate chart and the manifold are both nonsingular. Stated more generally, assigning a non-singular coordinate system to a four-manifold already introduces an orientation, allowing one to define integrals such as the action or measures of volume, area or length. Writing the metric as usual by $g_{\mu \nu}= e_\mu^A e_\nu^B \eta_{AB}$, with  $e_\mu^A$ the frame field and $\eta_{AB}$ the Minkowski metric, only one continuously connected component of non-singular frame fields $e_\mu^A$ -- for example the component with strictly positive eigenvalues -- is needed in order to describe a general, nonsingular four-geometry. To sum over additional components (for example to sum over both positive and negative lapse functions $N$ while taking the determinant $h$ to be positive) is not only unnecessary, it represents an overcounting which is unjustified from a geometrical point of view. Furthermore, although arbitrarily small $N$ should be allowed, one should {\it not} include the point $N=0$ in the sum since it does not describe a four-geometry. Finally, integrating over all Lorentzian four-geometries requires only {\it real} (and positive) values of $N$.  If that fundamental, geometrical definition can be deformed into a mathematically equivalent integral over complex metrics which is easier to calculate, as Picard-Lefschetz theory and Cauchy's theorem allow, that is all well and good. But it makes little geometrical sense to take an integral over complex lapse functions $N$ as a fundamental definition of the theory. 

In their most recent paper,  Diaz Dorronsoro  {\it et al.}~\cite{DiazDorronsoro:2018wro} misrepresent our work by stating that we ``have recently advanced a larger class of wavefunctions that extend the original" no-boundary wavefunction. Quite to the contrary, what we explained in our earlier papers is that the integral over Lorentzian four-geometries is actually {\it unique}! This allowed us to compute {\it the only} geometrically meaningful ``no boundary wavefunction." The fact that calculation failed to give an observationally acceptable result is not the fault of the path integral for gravity, but rather that of imposing the ``no boundary" idea in this particular model, attempting to describe the beginning of the universe in the context of inflationary scenarios.

In fact, it is Diaz Dorronsoro {\it et al.}, not us, who are ``advancing a larger class of wavefunctions" in an attempt to rescue the no-boundary proposal. As we have explained, there is no geometrical justification for taking an integral over complex metrics as a starting point for the theory. Yet this is exactly what they propose~\cite{DiazDorronsoro:2018wro}.  They consider metrics of the axial Bianchi IX form
\begin{align} \label{metric}
ds^2 = - \frac{N^2}{q}dt^2 + \frac{p}{4} (\sigma_1^2+\sigma_2^2) + \frac{q}{4} \sigma_3^2\,,
\end{align}
where $p(t),q(t)$ are time dependent scale factors and $\sigma_1 = \sin\psi d\theta - \cos \psi \sin \theta d\varphi$, $\sigma_2 = \cos \psi d\theta + \sin \psi \sin \theta d \varphi$, and $\sigma_3 = - (d\psi + \cos\theta d\varphi)$ are differential forms on the three sphere with $0 \leq \psi \leq 4 \pi$, $0 \leq \theta \leq \pi$, and $0 \leq \varphi \leq 2 \pi.$ For real $N>0$, this metric describes Bianchi IX spacetimes on the axes of symmetry. In the well-known notation of Misner \cite{Misner:1969hg} this corresponds to the line $\beta_-=0.$  Diaz Dorronsoro et al.\ now propose to define the gravitational path integral as a sum over real values of $p$ and $q,$ supplemented by a sum over values of the lapse function $N$, taken along a {\it complex} circular contour enclosing the origin.  

In our view this proposal is quite arbitrary, as it is not motivated by any fundamental physical principle. What does it mean to integrate over metrics with complex proper time intervals? In \cite{DiazDorronsoro:2018wro}, this sum over specific complex metrics is regarded not merely as a calculational device, but as the starting definition of the theory. Furthermore, this definition seems context dependent. Such a definition will neither allow one to calculate meaningful transition amplitudes between two large three-geometries nor to understand how quantum field theory on curved space-time emerges when the scale factor evolves classically. Given its poor motivation, we find it unsurprising that this definition ultimately leads to mathematical and physical inconsistencies, as we shall explain in the remainder of this note. 

Before doing so, it may be useful to briefly comment on the relation of the path integral to the propagator and the Wheeler-DeWitt equation. For concreteness, consider a simple relativistic path integral such as is encountered for a relativistic particle. Formally, the starting point is the relativistic propagator,
\begin{align}
G[q_1,q_0]=\langle q_1| \int_{0^+}^\infty \mathrm{d}N e^{-i \hat{\mathcal H} N} |q_0\rangle,
\label{prop1}
\end{align}
where $\hat{\mathcal H}$ is the Hamiltonian, where the rhs can be expressed as a path integral in the usual way. Applying the operator $\hat{\mathcal H} $ and rewriting $\hat{\mathcal H} e^{-i \hat{\mathcal H} N} =i (\mathrm{d} e^{-i \hat{\mathcal H} N}/\mathrm{d}N)$, the $N$ integral becomes a boundary term at $N=\infty$, which may be taken to vanish, minus another at $N=0^+$, proportional to $\langle q_1|q_0\rangle=\delta(q_1-q_0)$. In this way one obtains
\begin{align}
\hat{\mathcal H} G[q_1,q_0]= -i \delta(q_1-q_0). 
\end{align}
The proposal of \cite{DiazDorronsoro:2018wro} is to instead obtain a homogeneous solution of the Wheeler-DeWitt equation $\hat{\mathcal H} \Psi=0$ from a similar formula
\begin{align}
\Psi[q_1]=\langle q_1| \int_{\cal C} \mathrm{d}N e^{-i \hat{\mathcal H} N} |\chi\rangle,
\label{pstate}
\end{align}
where $\chi$ is any state, and ${\cal C}$ is a contour which yields no endpoint contributions. For example, ${\cal C}$ may start and end at infinity, or it may be closed. Whereas the propagator (\ref{prop1}) is uniquely defined,  (\ref{pstate}) in principle depends both on the state $|\chi\rangle$ and the contour ${\cal C}$. This infinite ambiguity is related to the fact that there are infinitely many homogeneous solutions of the Wheeler-DeWitt equation. In order to define a ``wavefunction of the universe," some other physical or mathematical principles are needed. In the model they study, Dorronsoro {\it et al.}\  take for ${\cal C}$  a small contour in the complex $N$-plane enclosing the origin. As already noted, there is little justification for this choice. Furthermore, it immediately leads to a problem with the path integral. Since there is no singularity in $N$ in the integrand of (\ref{pstate}), at fixed $p(t)$ and $q(t)$,  there is no obstruction to shrinking the $N$ contour away. This means that if one performs the $N$ integral first, the answer is zero! Diaz Dorronsoro {\it et al.}\ do not notice this because they perform the path integrals over $p(t)$ and $q(t)$ first, generating a pole in $N$ from the corresponding prefactors. Then the $N$ integral, taken on a closed contour enclosing the origin, extracts the residue. Clearly, their result depends on the order in which the partial integrals of the path integral are taken. While Dorronsoro {\it et al.}\  {\it do} generate a solution to the Wheeler-DeWitt equation this way, any connection to the original path integral is clearly on shaky ground. There are additional (and related) problems with their definition to which we return at the end of the next section.

\section{Normalizability} 
\label{section:norm}
With the metric \eqref{metric}, the action for gravity plus a cosmological constant $\Lambda,$ in units where $8\pi G =1,$ is given by
\begin{align}
S/(2\pi^2) = \int_0^1 \mathrm{d}t \left[ -\frac{1}{4N}\left(\frac{q \dot{p}^2}{p} + 2 \dot{p}\dot{q}\right) + N \left(4-\frac{q}{p} -p \Lambda\right) \right]\,,
\label{act1}
\end{align}
where the integrals over angular directions yield a factor of $16\pi^2$. In this section we evaluate the classical action. Then we apply Picard-Lefschetz theory to identify the relevant saddles and deform the $N$ integral to render it absolutely convergent. We also discuss the normalizability of the resulting ``wavefunction".

\subsection{The classical action}
The equations of motion corresponding to the variations of $q$ and $p$ are given by
\begin{equation}
2p\ddot{p} - \dot{p}^2  = 4N^2 \,,\qquad 
\ddot{q} + \frac{\dot{p}}{p}\dot{q}  = N^2 \left(2 \Lambda -\frac{4q}{p^2}\right)\,,
\label{second}
\end{equation}
and the constraint following from the variation of $N$ is given by
\begin{align}
\frac{1}{4}\left(\frac{q}{p}\dot{p}^2 + 2 \dot{p}\dot{q}\right) + N^2 \left(4-\frac{q}{p} -p \Lambda\right) =0\,.
\end{align}
Regular solutions to these equations, behaving as $p(t)\sim q(t)\sim \pm 2i N t$ as $t\rightarrow 0$, correspond to (part of) Taub-NUT-de Sitter spacetime. The corresponding complex, regular geometries are considered to be of the no-boundary type~\cite{DiazDorronsoro:2018wro}. We focus on these in what follows.

Since the Lagrangian in (\ref{act1}) is linear in $q(t)$, the path integral over $q(t)$ enforces a functional delta function for first equation of motion in (\ref{second}). Therefore the only paths $p(t)$ which contribute to the path integral are those which satisfy this equation. Using it, the action reduces to:
\begin{align}
S/(2\pi^2) &= \int_0^1 \mathrm{d}t \left( -\frac{1}{2N} \frac{\mathrm{d}}{\mathrm{d}t}\left(\dot{p}q\right) + N \left(4 -p \Lambda\right)  \right), 
\label{eq:Classical1}
\end{align}
so that the classical action depends on $q(t)$ only through its boundary values. 

In order to implement the no-boundary proposal, as explained above we take $p(0)=q(0)=0$. We also set $p(1)=p_1$, $q(1)=q_1$ where $p_1$ and $q_1$ are arbitrary positive constants, to describe the final, anisotropic three-geometry. With these boundary conditions, the equation of motion for $p$ has no real solution. There is, however, a pair of complex conjugate solutions, 
\begin{align}
p_\pm(t) = \pm i N t(t-1)+p_1 t^2.
\label{psol}
\end{align}
for which the corresponding classical action (\ref{eq:Classical1}) is given by 
\begin{align} 
S_{\pm}(N)/(2\pi^2) & = -\frac{p_1 q_1}{N} \pm i q_1  + N \left(4 - \frac{\Lambda}{3}p_1\right)  \mp i  \frac{\Lambda}{3} N^2 \,. \label{finalactionconv}
\end{align} 
We claim that the original, Lorentzian path integral over positive real values of $N$ is convergent. According to (\ref{finalactionconv}), after integrating out $p(t)$ and $q(t)$, for the two possible classical solutions (\ref{psol}), the semiclassical exponent $iS_\pm(N) \sim \pm  \frac{\Lambda}{3} N^2$ at large $N$. In order for the $N$ integral to converge, we must take $S_-(N)$, corresponding to the solution $p_-(t)$. This choice is actually in conflict with the ``momentum constraint'' imposed in \cite{DiazDorronsoro:2018wro} (in fact it corresponds to the opposite ``momentum constraint''), but it is mandatory if one starts from the Lorentzian path integral. Nevertheless, we will also later consider their choice of solution, $p_+(t)$ with action $S_+$, in order to highlight some aspects of this choice. But we emphasize that it is incompatible with the Lorentzian path integral.

\begin{figure}[h] 
\begin{center}
\includegraphics[width=0.6\textwidth]{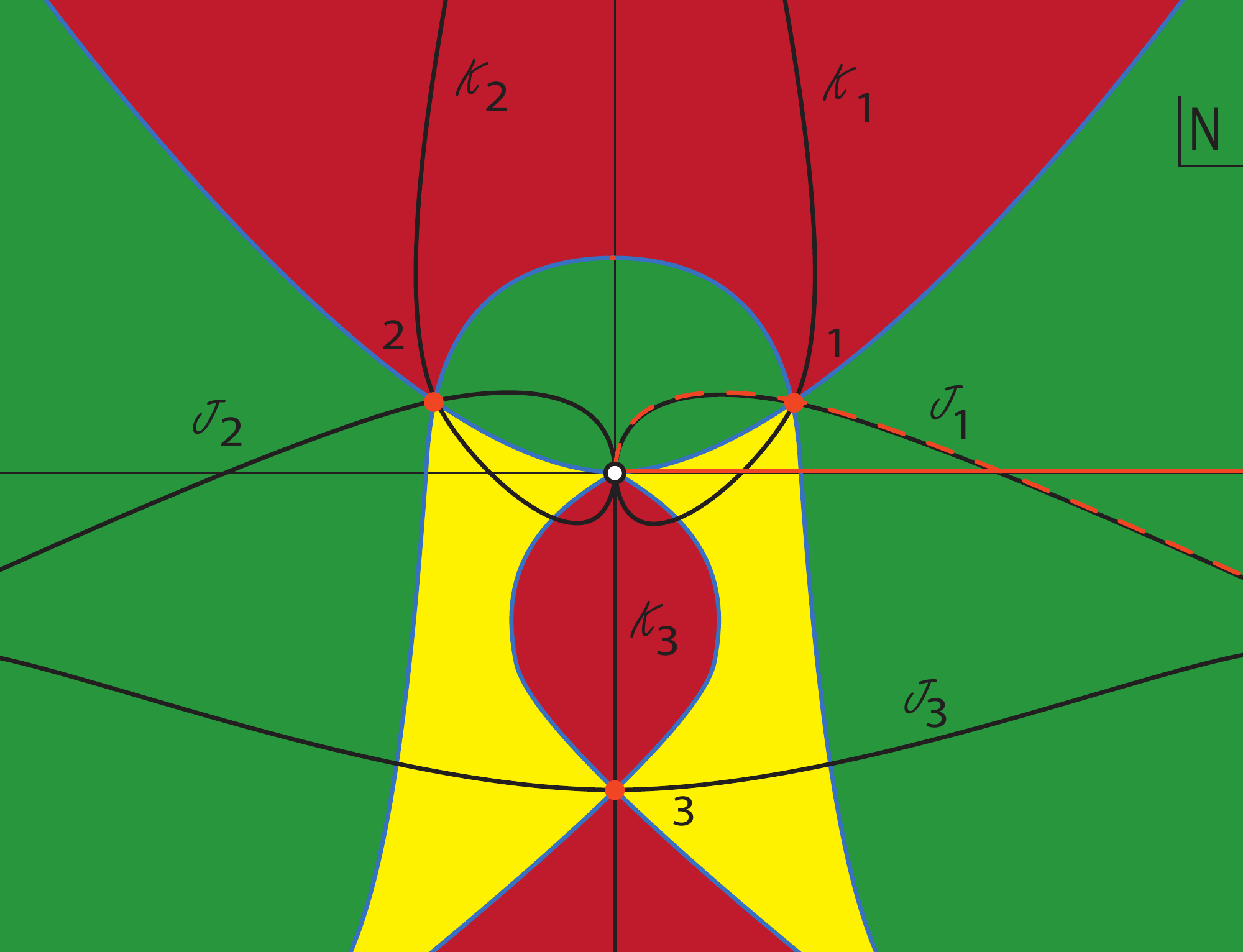}
\caption{The location of the saddle points and flow lines for the action we advocate, $S_-$ in \eqref{finalactionconv}, for which the Lorentzian integral is convergent. The saddle points are indicated by the orange dots. Green regions have a lower magnitude of the integrand than at the adjacent saddle point, red regions have a higher magnitude and yellow regions have a magnitude in between two saddle point values. If $N$ approaches the singular point at infinity or the essential singularity at $N=0$ along a contour in a green region, we obtain  a convergent integral. Conversely, if $N$ approaches these points along a contour in a red region, the integral diverges. }
\label{Fig:con}
\end{center}
\end{figure}

The corresponding propagators simplify to an oscillatory integral over the lapse, i.e.
\begin{equation}
G_\pm[q_1,p_1;0,0] \propto \int \frac{\mathrm{d}N}{N} e^{i S_\pm[q_1,p_1;0,0;N]/\hbar}\,.
\end{equation}

\begin{figure}[h] 
\begin{center}
\includegraphics[width=0.6\textwidth]{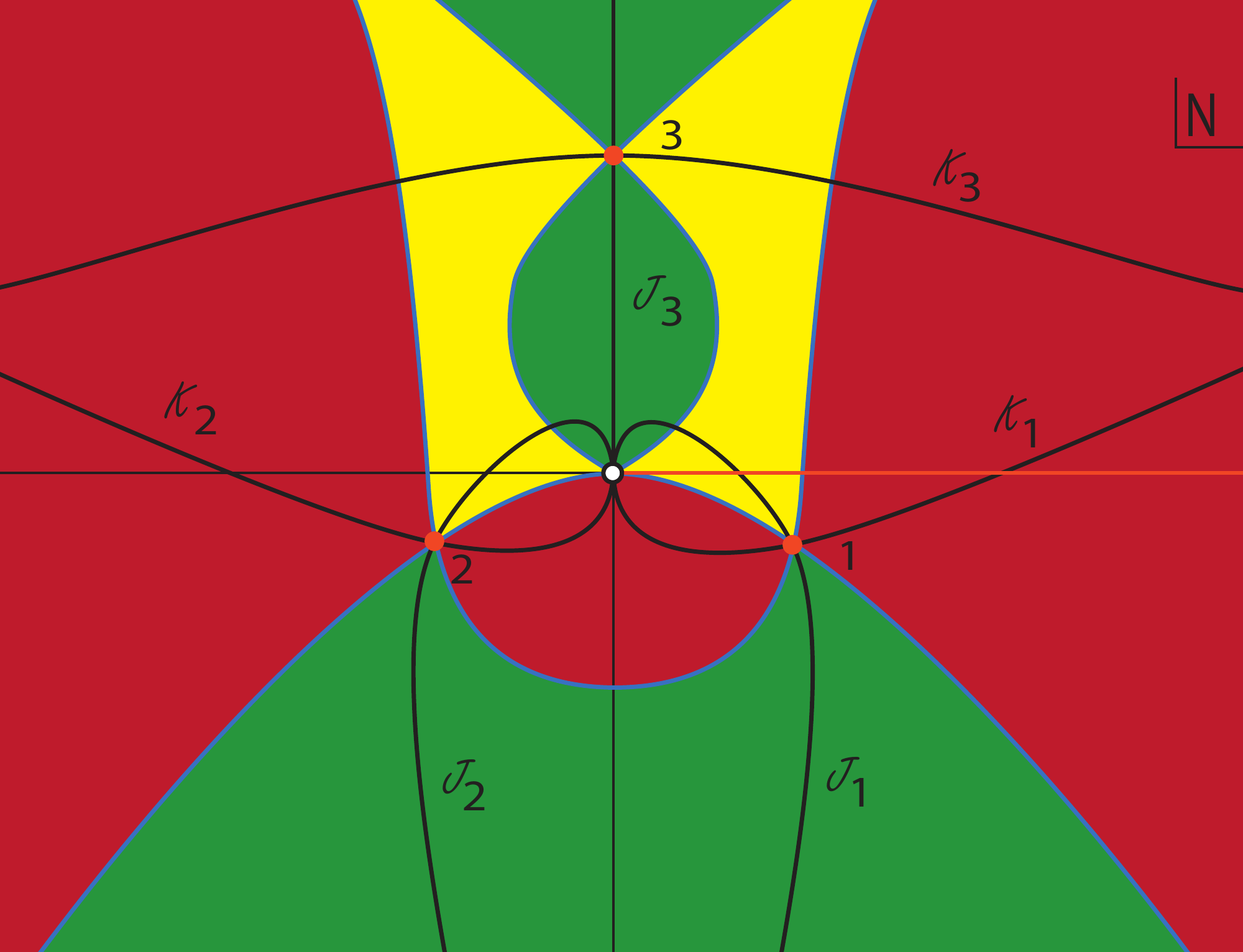}
\caption{The location of the saddle points and flow lines for the action $S_+$, for which the Lorentzian path integral diverges, but which is chosen by Diaz Dorronsoro  {\it et al.}~\cite{DiazDorronsoro:2018wro}. For a description of the colour scheme, see the caption of Fig. \ref{Fig:con}. Note that, as this Figure shows, it would also have been possible to define a purely Euclidean contour along the positive imaginary axis for this choice of action, and this would have led to only saddle point $3$ contributing. This latter saddle point leads to a purely Euclidean geometry, without any classical Lorentzian evolution.}
\label{Fig:div}
\end{center}
\end{figure}

\subsection{Picard-Lefschetz theory}

\begin{figure}[h] 
\begin{center}
\includegraphics[width=0.5\textwidth]{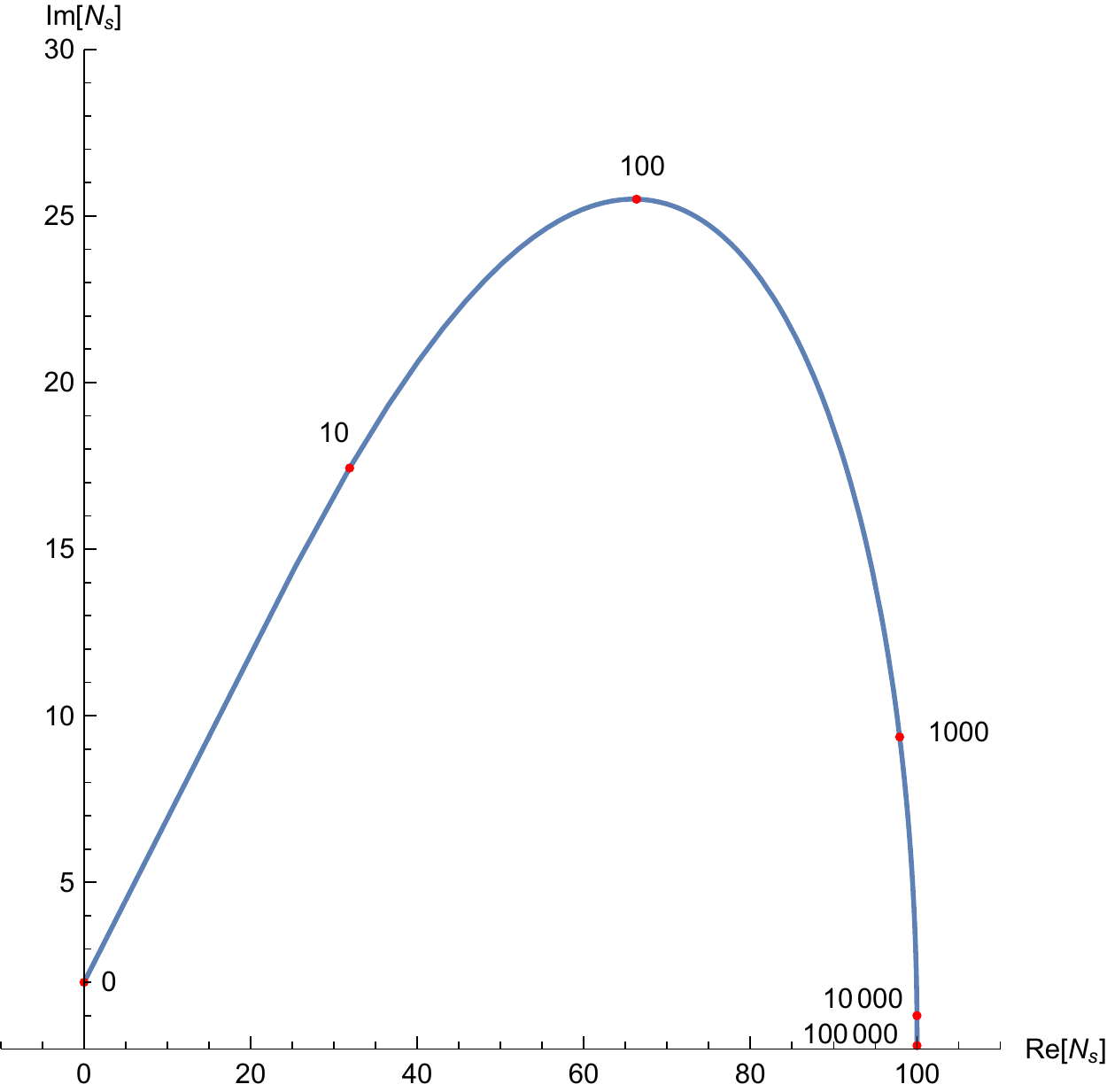}
\caption{The location of the relevant saddle point, for fixed $q_1=10000$ and as a function of $0 < p_1 < 100000$. Some indicative values of $p_1$ are shown next to the curve. At large values of $p_1$ the saddle point remains complex but moves very close to the real $N$ line.}
\label{AxialSaddle}
\end{center}
\end{figure}

\begin{figure}[h] 
\begin{minipage}{0.5\textwidth}
		\includegraphics[width=0.95\textwidth]{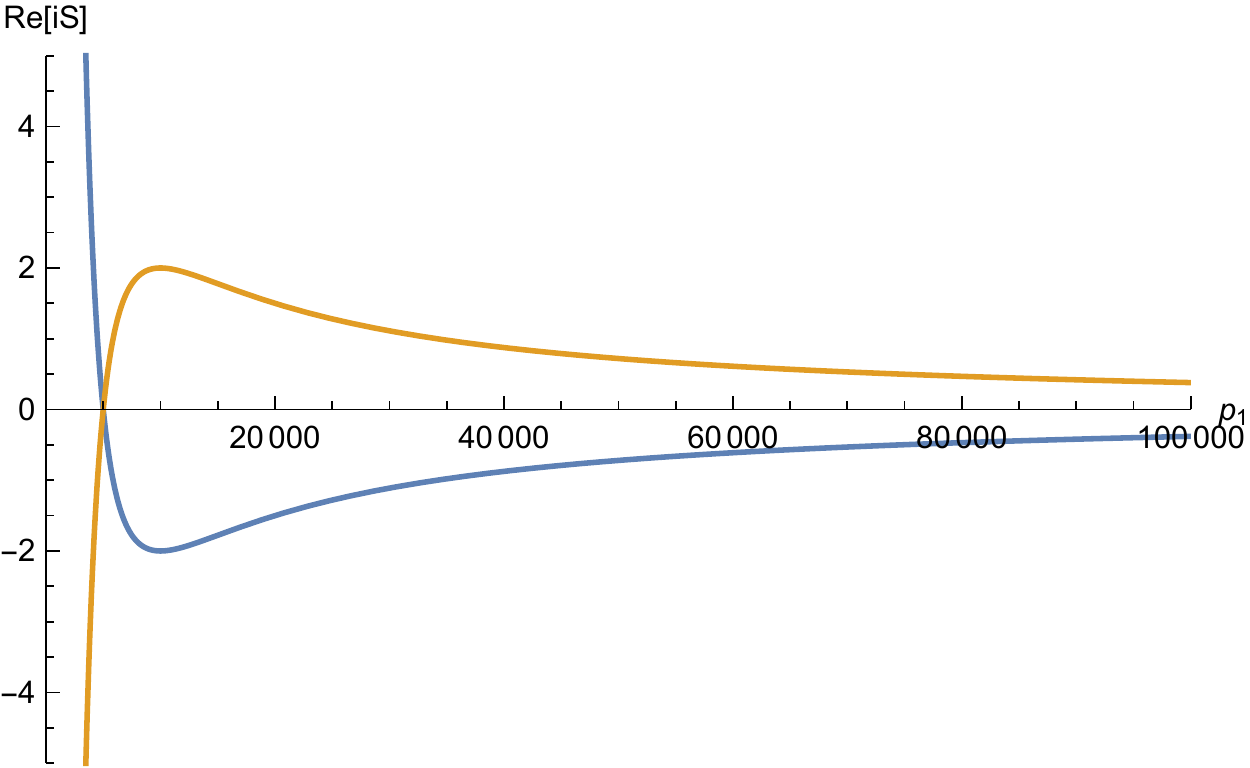}
	\end{minipage}%
	\begin{minipage}{0.5\textwidth}
		\includegraphics[width=0.95\textwidth]{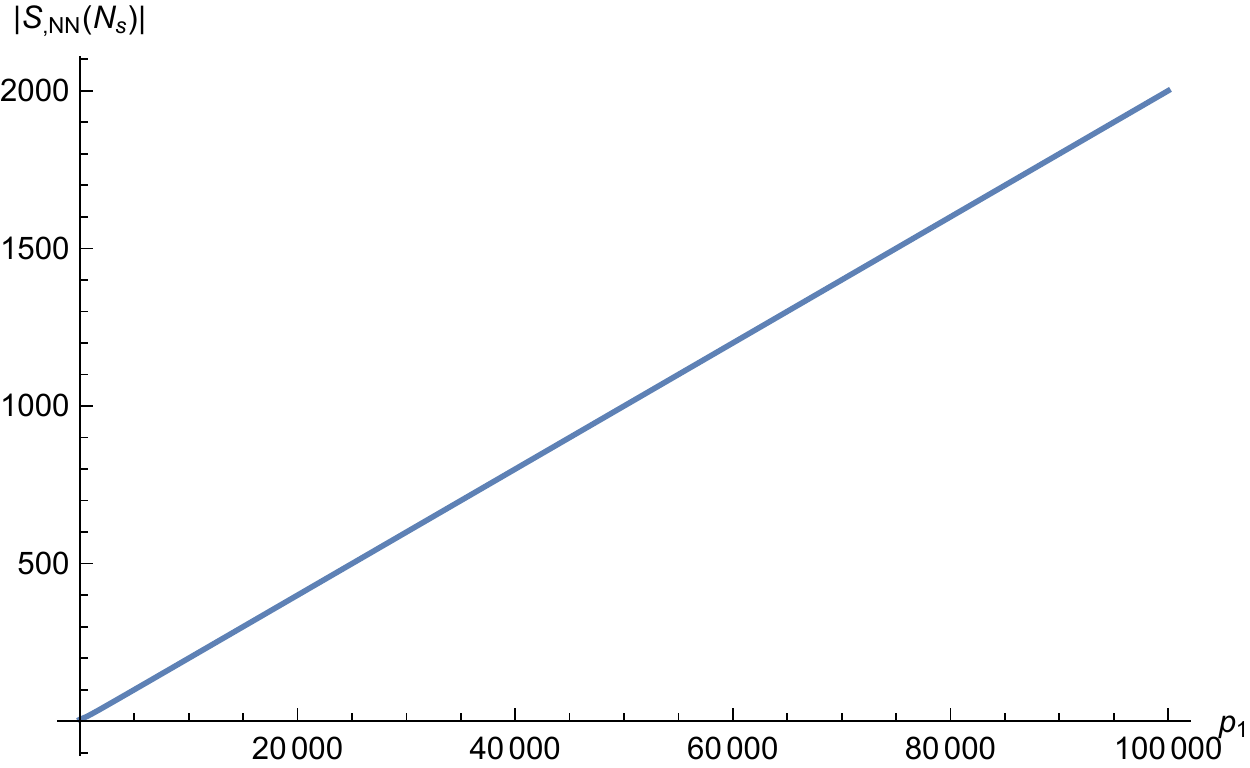}
	\end{minipage}%
\caption{{\it Left:} The Morse function $-Im(S)$ for fixed $q_1=10000$ and as a function of $0 < p_1 < 100000$ both for the relevant Lorentzian saddle point, with action $S_-(N)$, (blue) and for one of the saddle points advocated by Diaz Dorronsoro  {\it et al.}\ (orange), with action $S_+$. {\it Right:} The absolute value of the second derivative at the same saddle points for fixed $q_1=10000$ (with $\Lambda=3$) and as a function of $0 < p_1 < 100000.$ For large $p_1$ the saddle point approximation becomes better and better.}
\label{SecDer}
\end{figure}

Having reduced the path integral to an ordinary integral over the lapse function $N,$ we are now in a position to evaluate it in the saddle point approximation. Figures \ref{Fig:con} and \ref{Fig:div} show the locations of the saddle points and steepest ascent/descent lines emanating from them for the two choices of the action given in \eqref{finalactionconv}. It is straightforward to see that the integral over real Lorentzian metrics, with semiclassical action $S_-$,  can be deformed into the steepest descent contour ${\cal J}_1$ passing through saddle point $1.$ The location of this saddle point for various values of $p_1$ and fixed $q_1$ is shown in Fig. \ref{AxialSaddle}. For large anisotropies it moves closer and closer to the real $N$ line, without however ever reaching it. The induced weighting is shown by the blue curve of the left panel in Fig. \ref{SecDer}, where it can be seen that the isotropic boundary conditions (here $p_1=q_1=10000$) receive the lowest weighting. In other words, the model is out of control, as more anisotropic geometries are favoured. 

An even more dramatic failure of the model is seen by sending $q_1$ to large values. The second term in $S_-$, given in \eqref{finalactionconv}, contributes a semiclassical exponent $+q_1$ which clearly leads to a non-normalizable wavefunction. Writing the volume of the final three-geometry as $V\sim p_1 q_1^{1\over 2}$ (see (\ref{metric})), and the anisotropy as $\alpha =q_1/p_1$, the terms which determine the saddle point value of $N$ in $S_-(N)$ at large $N$ (see \eqref{finalactionconv}) are the first and the last. Evaluating the action at this saddle, one finds the second term dominates for anisotropy $\alpha > V^{1\over 2} \Lambda^{3\over 4}$. Defining the de Sitter radius $R_{H}\equiv \Lambda^{-{1\over 2}}$, one sees that the semiclassical exponent $i S_-$ is dominated by the $+q_1$ term as long as the length in the 3 direction (see (\ref{metric})), $ q_1^{1\over 2} = \alpha^{1\over 3} V^{1\over 3}$ exceeds $V^{1\over 2} R_{H}^{-{1\over 2}}$ while the length in the other two spatial directions $p_1^{1\over 2}=V^{1\over 3}/\alpha^{1\over 6}$ is smaller than $V^{1\over 4} R_{H}^{1\over 4}$. Clearly, both lengths can grow without bound as $V$ is increased while remaining in this regime. So the semiclassical exponent can become arbitrarily large, whilst all curvature scales remain greater than the Planck length so that semi-classical Einstein gravity remains valid. 

These findings, that imposing ``no boundary" boundary conditions in the Lorentzian path integral lead to an unacceptable amplitude, favouring large deviations from isotropy,  confirm those of our earlier analysis of inhomogeneous perturbations around the background FLRW cosmology~\cite{Feldbrugge:2017fcc,Feldbrugge:2017mbc}. The advantage of the present discussion is that it is fully non-perturbative, albeit still semi-classical. Hence, our analysis removes any hope that a treatment going beyond cosmological perturbation theory might yet rescue the no-boundary proposal. As explained previously, we adhere strictly to the Lorentzian formulation of the gravitational path integral which, we have argued, is the only one with a chance of making mathematical and physical sense.

As shown by Diaz Dorronsoro  {\it et al.}, if one takes the integral over $N$ along a circular contour around $N=0$ for the action $S_+$ in \eqref{finalactionconv}, the contour can be deformed to a sum over the two steepest descent paths ${\cal J}_1$ and ${\cal J}_2$ in Fig.~\ref{Fig:div}. These saddle points lie respectively at the complex conjugate and negative values of the Lorentzian saddle point $1$ in Fig.~\ref{Fig:con}.
%(and whose asymptotic location at large $p_1$ is given by \eqref{saddlelargep1}). 
The weighting of these saddle points is just the inverse of the weighting of the Lorentzian saddle point, and is shown by the orange curve in the left panel of Fig.~\ref{SecDer}. For these the isotropic configuration $p_1=q_1$ is indeed the configuration with the highest weighting. However, having a maximum is not enough to ensure normalizability. Indeed, just as for the Lorentzian saddle point, the weighting of these saddle points tends to a constant at large values of $p_1$ (the inverse of a constant being another constant), so that again an integral of the weighting $e^{-2 Im(S_{+})/\hbar}$ over $p_1$ is unbounded, and the corresponding wavefunction is non-normalizable. Thus if normalizability is regarded as a crucial criterion, the new circular contour must also be discarded on these grounds. 

For reasons that are not clear to us, the authors of \cite{DiazDorronsoro:2018wro}, even though they also noticed the unboundedness of the integral, simply chose to truncate it by hand. The stated reason was that the approximations involved in the calculation break down. However, this statement is puzzling, as the axial Bianchi IX model allows one to calculate the action exactly and, moreover, the saddle point approximation becomes better and better at large $p_1$ (see again the right panel of Fig. \ref{SecDer}, which also applies to the saddle points in question). Thus the implied non-normalizability seems robust, to the extent that normalizability is understood at all. 

\section{Mathematical and Physical Consistency}

We now come to what we regard as the biggest flaw in the proposal of Diaz Dorronsoro  {\it et al.}, namely that it seems to us to fail some simple tests of physical and mathematical consistency. When we take the limit in which the final three-geometry is isotropic, it seems reasonable to expect that we should recover the result of the truncated, isotropic theory, at least in the semi-classical limit where quantum backreaction is negligible. Likewise, if we add an additional metric perturbation mode to the final three-geometry, for example one of an inaccessibly small wavelength, this should not immediately lead to inconsistent results. We will discuss these two tests of their proposal, in turn.

\begin{figure}[h] 
\begin{center}
\includegraphics[width=0.5\textwidth]{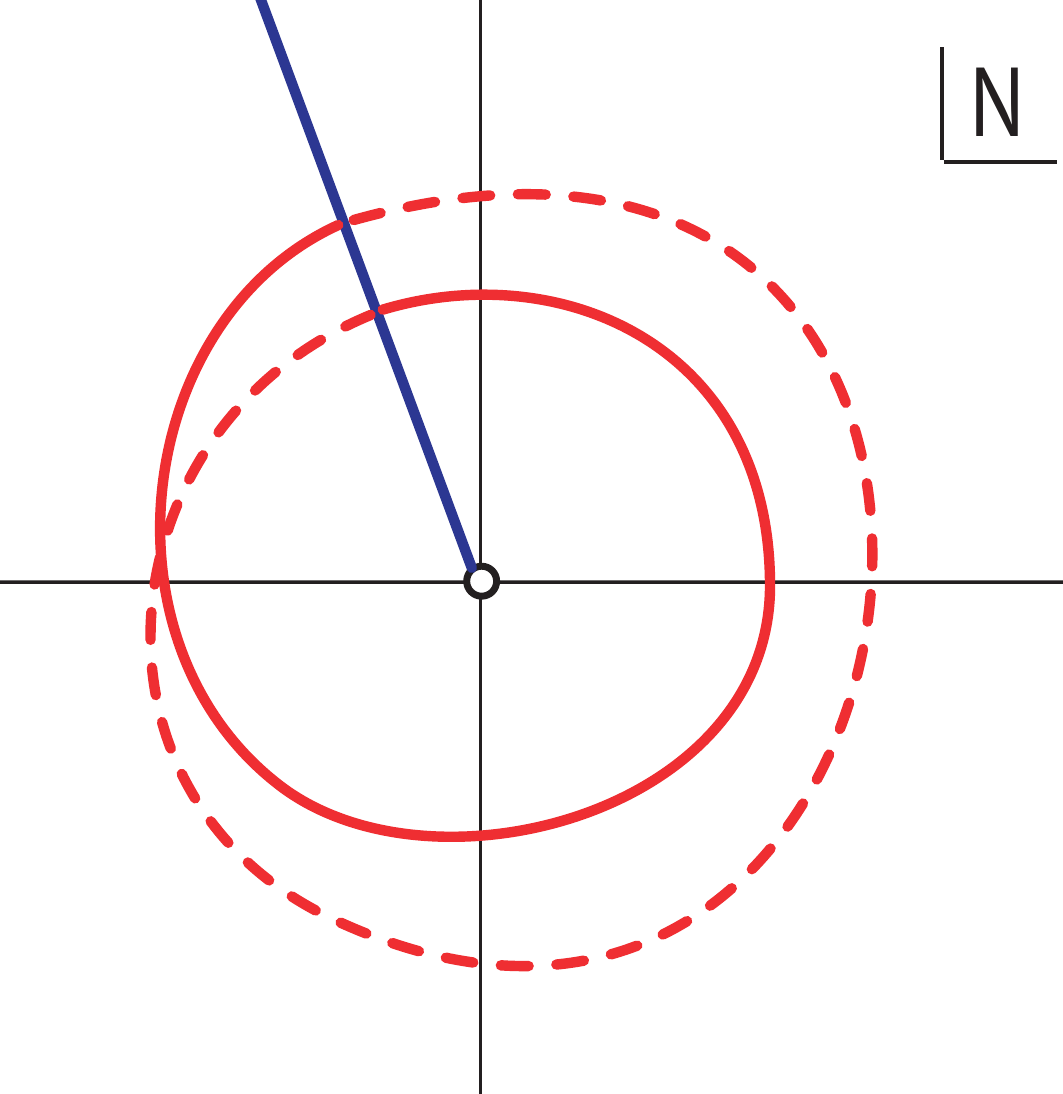}
\caption{An example of a circular contour in the presence of a branch cut (in blue), for a two-sheeted integrand. The contour must complete two loops before it can close.}
\label{Fig:circular}
\end{center}
\end{figure}

First, consider the isotropic limit, where $p_1=q_1.$ Here we would expect the axial Bianchi IX model to reproduce the results of an isotropic FLRW minisuperspace model, defined using the same integration contour for the lapse function. Certainly, for Lorentzian integrals, this is the case and the action $S_-$ in  Eq. \eqref{finalactionconv} indeed reproduces our earlier isotropic results of \cite{Feldbrugge:2017kzv}. When $p_1=q_1,$ the relevant saddle point of the action $S_-$ is located at
\begin{align}
N_{s1}^{iso}=\sqrt{\frac{3}{\Lambda}}\sqrt{q_1-\frac{3}{\Lambda}} + i \frac{3}{\Lambda}\,, \label{isosaddle}
\end{align}
{\it i.e.}, it resides at the same value of $N$ as for the isotropic model, where the action is given by a different function of $N,$ namely \cite{Halliwell:1988ik,Feldbrugge:2017kzv}
\begin{align}
S^{iso}(N)/2\pi^2=\left[ N^3 \frac{\Lambda^2}{36} + N\left(3 - \frac{\Lambda}{2}q_1\right) -\frac{3q_1^2}{4 N} \right]\,. \label{isoaction}
\end{align}
Moreover, at the isotropic saddle point \eqref{isosaddle}, the values of the axial Bianchi IX action $S_-$ and the isotropic action \eqref{isoaction} agree,
\begin{align}
S_{conv}(N_{s1}^{iso}) = S^{iso}(N_{s1}^{iso}) = 2\pi^2 \left( -\frac{2\sqrt{3}}{\Lambda}(\Lambda q_1 - 3)^{3/2} + i \frac{6}{\Lambda}\right)\,.
\end{align}
Thus we find a well-behaved isotropic limit, as we believe we should, since in the isotropic limit we are describing the same physical situation.

However, when we take the circular contour advocated by Diaz Dorronsoro  {\it et al.}, a problem arises. In the isotropic case, the path integral reduces to an ordinary integral over the lapse function of the form \cite{Halliwell:1988ik,Feldbrugge:2017kzv}
\begin{align}
G[q_1,0]=\sqrt{\frac{3 \pi i}{2 \hbar}} \int  \frac{\mathrm{d}N}{\sqrt{N}} e^{i S^{iso}(N)/\hbar}\,.
\end{align}
The prefactor, which arises from the integral over the isotropic scale factor, contains a factor of $1/\sqrt{N},$ so that there is a branch cut in the integrand, emanating from the origin. This branch cut requires that a circular contour must complete two loops around the origin before it can close -- see Fig. \ref{Fig:circular}. However, on the second loop the factor $1/\sqrt{N}$ will acquire a minus sign relative to its value on the first loop, so that the contributions from the second loop exactly cancel those of the first loop. The result is that, for isotropic boundary conditions, a closed circular contour yields precisely zero! Hence there is blatant disagreement with the isotropic limit of the Bianchi IX model, although the physical situation being described is identical. (One may easily verify that the saddle points contributing to the path integral with final boundary $p_1=q_1$ also have $p(t)=q(t)$ throughout the entire geometry $0 \leq t \leq 1$). Hence this choice of contour fails to satisfy our consistency check. 

The second inconsistency manifests itself in the following, closely related manner. In minisuperspace models, when we include $n$ deformations of the metric in addition to the lapse,  the prefactor generally takes the form $1/N^{n/2}$~\cite{Halliwell:1988ik}. For $n$ odd, the integrand will thus be taken around a branch point at $N=0$ and a closed contour about the origin will again yield a vanishing result. But the results of our calculations should not depend on how many possible deformations we include as long as the same physical situation is described. One should be able to add a possible deformation and then consider boundary conditions in which this additional deformation is zero -- and the results should, at this leading semi-classical level, be unchanged. A straightforward example is to use the full Bianchi IX metric and then restrict to boundary conditions corresponding to the axial Bianchi IX truncation studied in this paper. Once again this does not lead to consistent results, as the Bianchi IX metric contains one additional deformation, so that a closed contour enclosing the origin again leads to a vanishing wavefunction.

\section{Discussion}

When constructing theories of the very early universe, the difficulty of making direct observations means that mathematical and physical consistency requirements must necessarily play a critical, guiding role. In our view, the new path integral for semi-classical gravity advocated by Diaz Dorronsoro  {\it et al.}~\cite{DiazDorronsoro:2018wro}, involving a closed integral for the complexified lapse function, seems inadequate in this regard: it has no geometrical interpretation as it involves metrics with complex proper times. Likewise, it abandons any notion of causality from the outset. Furthermore, when describing the same physical situation using different truncations of the degrees of freedom in the spacetime metric, it yields vastly different results. The clearest example is provided by truncating the model to an isotropic, one-dimensional minisuperspace, for which a closed contour about the origin yields a vanishing ``wavefunction." More generally, such a closed contour fails to yield a meaningful wavefunction for any odd-dimensional truncation of minisuperspace -- violating the physically reasonable requirement that including one additional mode, for example one with an inaccessibly tiny wavelength, should not change a physical result. A general formal argument against ad hoc contours of the type Diaz Dorronsoro  {\it et al.}\ consider was given at the end of Section II. Unless a closed contour starts and ends at the point at infinity in the complex $N$-plane, it cannot give a nonzero result unless the path integral depends on the order in which the integrals are taken. Such a theory is clearly ambiguous at best. 

In previous work, we have shown that attempts to define a ``smooth beginning" for inflation based on either the no-boundary proposal \cite{Hartle:1983ai} or the tunneling proposal  \cite{Vilenkin:1982de,Vilenkin:1984wp,Vilenkin:1994rn} are either mathematically inconsistent or they lead to the physically unacceptable results. The semiclassical ``no boundary" path integral taken over real Euclidean metrics fails in the first manner, whereas the path integral taken over real Lorentzian metrics, as posited in the ``tunneling proposal," fails in the second since it favors wildly fluctuating geometries. (In appendix \ref{AppB} we show that the recent rescue of the tunneling proposal proposed by Vilenkin and Yamada \cite{Vilenkin:2018dch}) unfortunately fails due to the breakdown of perturbation theory, {\it i.e.}, a strong coupling problem.)  At the root of this disaster are two key assumptions; i) that it makes sense to compute an amplitude for an ``out" state when there is no ``in" state (or when the ``in" state is replaced by a ``three-geometry of zero size") and ii) that the universe started out dominated by some kind of inflationary energy, which behaves in effect like a large, temporary cosmological constant. It is not yet clear to us whether abandoning one of these assumptions would lead to a more acceptable result. Conceivably, one should abandon both (see, {\it e.g.}, \cite{Gielen:2015uaa, Gielen:2016fdb}). 

As well as these negative conclusions, our investigations have also opened up a very interesting avenue to pursue. We have found many indications that the Lorentzian path integral for gravity, tackled consistently using Picard-Lefschetz theory and with sensible  ``in" and ``out" states, has a remarkable physical and mathematical uniqueness and consistency~\cite{Feldbrugge:2017kzv,Feldbrugge:2017fcc,Feldbrugge:2017mbc}. In future work, we shall outline what we consider to be a far less arbitrary and more promising approach to the problem of the initial conditions for the universe, based on a precise treatment of this gravitational path integral. 

%%%%%%%%%%%%%%%%%%%%%%%%%%%%%%%%%%%%

%%%%%%%%%%%%%%%%%%%%%%%%%%%%%%%%%%%%er

\acknowledgments

We sincerely thank Sebastian Bramberger, Angelika Fertig, Laura Sberna and Alice Di Tucci for collaboration on these topics. We also thank Juan Diaz Dorronsoro, Jonathan Halliwell, Alex Vilenkin, Yannick Vreys and Masaki Yamada for helpful discussions and correspondence. Research at Perimeter Institute is supported by the Government of Canada through Innovation, Science and Economic Development, Canada and by the Province of Ontario through the Ministry of Economic Development, Job Creation and Trade. JLL gratefully acknowledges the support of the European Research Council in the form of the ERC Consolidator Grant CoG 772295 ``Qosmology''.

%\newpage
\appendix
\section{Conditionally convergent integrals} \label{AppA}

Highly oscillatory integrals, such as the Fresnel integrals, play an important role in physics. They are of particular interest in Lorentzian quantum physics, as they arise naturally in the form of real-time path integrals. In minisuperspace models of Lorentzian quantum cosmology, the integral over the lapse function is quite generically of this type. In this appendix we will spell out a simple proof that such integrals, defined over real values of the lapse, are convergent, using only real field values and without requiring a complexification as is used in Picard-Lefschetz theory. Our proof will simply demonstrate the convergence without however showing what value the integral converges to. Then, as used earlier in the paper, one may conveniently use Picard-Lefschetz theory to obtain a saddle point approximation to the integral.

The integral of the function $g$ over an infinite domain is defined as
\begin{align}
\int_0^\infty \mathrm{d} x\ g(x) = \lim_{R\to \infty}\int_0^R \mathrm{d} x\ g(x)\,.
\label{eq:improper}
\end{align}
Assuming that the integral is convergent, the integral is called absolutely convergent when the integral over the magnitude of the integral converges
\begin{equation}
\int_0^\infty \mathrm{d} x\ |g(x) | < \infty\,.
\end{equation}
The integral is called conditionally convergent when the integral over the magnitude diverges
\begin{equation}
\int_0^\infty \mathrm{d} x\ |g(x) | = \infty\,.
\end{equation}
Conditionally convergent integrals, converge due to cancellations from violent oscillations for large $x$. 
It is important to note that for conditionally convergent integrals, as for conditionally convergent series, the result depends on the order of summation. For integrals this is naturally prescribed by definition \eqref{eq:improper}. By changing the order of summation one can engineer the integral to converge to any number. However, for such a deformation of the theory one can no longer use complex continuations and deformations of the integration contour in the complex plane to evaluate these integrals. This would in particular invalidate the commonly used $i \epsilon$ and Wick rotation methods in quantum physics. For this reason we stick to the natural definition and argue that conditionally convergent integrals are well-defined in real time quantum physics.

For many oscillatory integrals, convergence can be demonstrated with the Leibniz convergence test for alternating series. A real alternating series is defined as
\begin{equation}
a=\pm \sum_{i = 0}^\infty (-1)^i a_i\,,
\end{equation}
with $a_i$ positive real numbers. The Leibniz convergence test states that the series is convergent when the arguments decrease monotonically, i.e. $a_{i+1} \leq a_i$ for sufficiently large $i$, and the argument goes to zero in the limit of large $i$, i.e. $\lim_{i\to \infty}a_i =0$. To see the relation to oscillatory integrals, consider the integral
\begin{equation}
I = \int_0^\infty \mathrm{d}x\ e^{i f(x)}\,,
\end{equation}
for a real valued polynomial $f$ (for the Fresnel integrals $f(x)=x^2$). The real and imaginary parts of $I$ are given by
\begin{align}
\text{Re}[I] = \int_0^\infty \mathrm{d}x \ \cos(f(x))\,,\\
\text{Im}[I] = \int_0^\infty \mathrm{d}x \ \sin(f(x))\,.
\end{align}
For simplicity we concentrate on the real part. Let us assume that the leading term of $f$ goes like $x^n$ in the limit $x \to \infty$ for $n \in \mathbb{N}$. A change of coordinates $u = x^n$ gives the integral
\begin{align}
\text{Re}[I] = \int_0^\infty \frac{\mathrm{d}u}{n u^{1-1/n}} \ \cos(f(\sqrt[n]{u}))\,,
\end{align}
and ensures that $f(\sqrt[n]{u}) \sim u$ for large $u$. Now let the zero crossings of the argument be given by $z_i$ for $i \in \mathbb{N}$. The real part of the integral can be written as an alternating series
\begin{align}
\text{Re}[I] &= \left[\int_0^{z_0} + \sum_{i=0}^\infty \int_{z_i}^{z_{i+1}} \right]\frac{\mathrm{d}u}{n u^{1-1/n}} \ \cos(f(u^{1/n}))\,,\\
&= c \pm \sum_{i=0}^\infty (-1)^i \left|\int_{z_i}^{z_{i+1}} \frac{\mathrm{d}u}{n u^{1-1/n}} \ \cos(f(u^{1/n}))\right|\,,\\
&= c \pm \sum_{i=0}^\infty (-1)^i a_i\,,
\end{align}
without changing the order of summation, with $c$ the integral over the interval $(0,z_0)$, either the positive or the negative sign for $\pm$ depending on the details of $f$, and the positive real numbers
\begin{equation}
a_i = \left|\int_{z_i}^{z_{i+1}} \frac{\mathrm{d}u}{n u^{1-1/n}} \ \cos(f(u^{1/n}))\right|\,.
\end{equation}
The argument of the alternating series can be dominated with a simple approximation
\begin{align}
a_i &= \left|\int_{z_i}^{z_{i+1}} \frac{\mathrm{d}u}{n u^{1-1/n}} \ \cos(f(u^{1/n}))\right|\\
&< \int_{z_i}^{z_{i+1}} \frac{\mathrm{d}u}{n u^{1-1/n}} = \sqrt[n]{ z_{i+1}}- \sqrt[n]{ z_{i}}=b_i\,.
\end{align}
In the limit of large $u$ the function $f(\sqrt[n]{u})$ asymptotes to a function proportional to $u$. For this reason, in the limit of large $i$, the zero crossings $z_i$ in $u$ will asymptote to a regular spacing, leading to the conclusion that for $n>1$ and for sufficiently large $i$, the coefficients $b_i$ satisfy the conditions of the Leibniz convergence test. Since $a_i < b_i$ for all $i$ we conclude that $\text{Re}[I]$ converges when $n > 1$. A similar argument can be given for the imaginary part of $I$, making $I$ conditionally convergent. 

The discussion above applies to a more general class of integrals. When the integral function $f(x)$ diverges as $x^{-n}$ in the limit $x \to 0$ with $n\in \mathbb{N}$, the change of coordinates $u = x^{-n}$ leads to convergence for $n>1$. More generally, when $f$ is not a polynomial but dominates some polynomial $x^n$ with $n>1$, the oscillatory integral can be shown to converge due to cancellations from oscillations at large $x$. Note that generally one should include the prefactor in the analysis.

The propagator for axial Bianchi IX consists of an oscillatory integral over the lapse $N$
\begin{equation}
G[q_1,p_1;q_0,p_0] \propto \int_0^\infty \frac{\mathrm{d}N}{N} e^{i S[q_1,p_1;q_0,p_0;N]/\hbar}\,,
\end{equation}
where the classical action $S[q_1,p_1;q_0,p_0;N] \sim N^2$ as $N\rightarrow \infty$ and $\sim N^{-1}$ for $N\rightarrow 0$. The discussion above directly shows that the integral converges at large $N$. The behavior of the integral for small $N$ is more subtle, since the discussion of the Leibniz convergence test is agnostic about polynomials for which $n=\pm 1$. By including the prefactor in the analysis, we now show that convergence is guaranteed. Consider the real integral
\begin{equation}
I = \int_0^1 \frac{e^{i/x}}{x}\mathrm{d}x\,.
\end{equation}
The integral does not converge absolutely. However, by a change of variables $u= -\ln x$, we can write the integral as
\begin{equation}
I = \int_0^\infty e^{i e^{u}}\mathrm{d}u\,.
\end{equation}
We treat the real and imaginary part of the integral separately and for simplicity concentrate on the real part. The real part can be written as an alternating series
\begin{align}
\text{Re}[I] &= \int_0^\infty \cos(e^u)\mathrm{d}u = \left[\int_0^{z_0} + \sum_{i=0}^\infty \int_{z_i}^{z_{i+1}} \right]\cos(e^u) \mathrm{d}u\\
&= c - \sum_{i=0}^{\infty}(-1)^i \left| \int_{z_i}^{z_{i+1}} \cos(e^u)\mathrm{d}u\right|\,,
\end{align}
where $z_i = \ln \left(\left(i + \frac{1}{2}\right)\pi\right)$ are the roots of the integrand. The arguments of the sum satisfies the Leibnitz condition since
\begin{equation}
\left| \int_{z_i}^{z_{i+1}} \cos(e^u)\mathrm{d}u\right| \leq z_{i+1}-z_i = \ln\left(\frac{2i+3}{2i+1 }\right) \,.
\end{equation}
Since the summands satisfy Leibniz condition for alternating series, the real part of the integral exists. Convergence of the imaginary part follows analogously. We thus conclude that the Lorentzian propagator for axial Bianchi IX is well defined as a conditionally convergent integral.

\section{Remark on the new tunneling proposal} \label{AppB}

The ``tunneling proposal" of Vilenkin \cite{Vilenkin:1982de} is closely related to the no-boundary proposal: it proposes to view the origin of the universe as a smooth tunneling event from ``nothing," defined as a three-geometry of ``zero size."  If one imposes that in the limit where the initial three-geometry is taken to zero size, the resulting semiclassical saddle point solution should be a regular complex solution of the Einstein equations, then there is no associated boundary term and the tunneling  proposal leads to exactly the same path integral as that we consider for the no-boundary proposal~\cite{Feldbrugge:2017kzv}. Consequently the tunneling proposal suffers from exactly the same instability. However, there is an ambiguity in the tunneling proposal's prescription of taking the initial three-geometry to have zero size: one could allow it to have large fluctuations in its local curvature or ``shape." In an attempt to rescue the  tunneling proposal, Vilenkin and Yamada have attempted to exploit this ambiguity by adding a boundary term for the fluctuations on the initial three-geometry, which they precisely tune in order to control the distribution for the fluctuations on the final three-geometry~\cite{Vilenkin:2018dch}. 

In their paper, they work in linear perturbation theory, where the fluctuations are regarded as small perturbations around the smooth, complex, classical background saddle point solution. The equation of motion for a Fourier mode of the perturbation $\phi_k$ (with wavenumber $k$) is of second order and it admits two solutions, which can be thought of as the positive and negative frequency modes, as usual. For instance, at the de Sitter saddle point, the general solution is given by a linear combination of two modes, \cite{Feldbrugge:2017fcc}
\begin{align}
\phi_k(\tau) = c_1 \frac{(1-\cos(H\tau))^{(k-1)/2}(\cos(H\tau)+k)}{(1+\cos(H\tau))^{(k+1)/2}} + c_2 \frac{(1+\cos(H\tau))^{(k-1)/2}(\cos(H\tau)-k)}{(1-\cos(H\tau))^{(k+1)/2}}\,,
\end{align}
which represent the analogue of the Bunch-Davies solutions for the closed slicing of de Sitter space. Here we have expressed the solution in terms of Euclidean time $\tau,$ so that the scale factor is given by $a= \frac{1}{H} \sin(H\tau).$ Near the origin $a=0$ (which also corresponds to the origin of $\tau,$ with $a \approx \tau$ for small $\tau$) these two modes behave very differently: the mode proportional to $c_1$ tends to zero as $\tau^{k-1}$, while the mode proportional to $c_2$ tends to infinity as $\tau^{-k-1}$. Since the action of the second mode diverges due to contributions near $a=0,$ we have discarded this mode in our previous works~\cite{Feldbrugge:2017mbc}. The mode proportional to $c_1$ leads to a finite action, but unfortunately it also leads to an inverse Gaussian distribution for the final perturbations. Vilenkin and Yamada propose to add a boundary term to the action, which has precisely the effect of cancelling the divergence of the action of the $c_2$ mode, while rendering that of the $c_1$ mode infinite. With their new proposal, one would then be led to pick out the stable $c_2$ mode. 

There is a good reason, however, for why this new proposal fails in its current form. The treatment of Vilenkin and Yamada rests on the applicability of a perturbative expansion around the classical background. In the presence of a perturbation mode such as the one above, the equation of motion for the background is corrected at quadratic order by terms involving the linear perturbations, namely (see e.g. \cite{DiTucci:2018fdg})
\begin{align}
- 2 \frac{a_{,\tau\tau}}{a} - \frac{a_{,\tau}^2}{a^2} + \frac{1}{a^2} = \Lambda + \frac{1}{2} \phi_{,\tau}^2 + \frac{(k^2 - 1)}{6 a^2} \phi^2 \label{back1}
\end{align}
In order for perturbation theory to be valid, the perturbative terms must be small compared to the background terms, in particular they must be small compared to the cosmological constant $\Lambda.$ But for the $c_2$ mode the perturbative terms scale as $\tau^{-2k-4}$ near $a=0.$ Thus at the ``bottom'' of the instanton  the solutions become entirely untrustworthy. Put differently, in going beyond the leading term in perturbation theory, one would encounter an infinity of additional terms, which would all blow up at small values of the scale factor, leading to a strong coupling problem where the theory is out of control. We thus conclude that, unfortunately, Vilenkin and Yamada's proposed rescue of the tunneling proposal also fails. 

\bibliographystyle{utphys}
\bibliography{Inconsistencies}

\providecommand{\href}[2]{#2}\begingroup\raggedright\begin{thebibliography}{10}

\bibitem{DiazDorronsoro:2018wro}
J.~Diaz~Dorronsoro, J.~J. Halliwell, J.~B. Hartle, T.~Hertog, O.~Janssen, and
  Y.~Vreys, ``{Damped perturbations in the no-boundary state},''
\href{http://arxiv.org/abs/1804.01102}{{\tt arXiv:1804.01102 [gr-qc]}}.
%%CITATION = ARXIV:1804.01102;%%.

\bibitem{Hartle:1983ai}
J.~B. Hartle and S.~W. Hawking, ``{Wave Function of the Universe},''
\href{http://dx.doi.org/10.1103/PhysRevD.28.2960}{{\em Phys. Rev.} {\bf D28}
  (1983)  2960--2975}.
%%CITATION = PHRVA,D28,2960;%%.

\bibitem{Feldbrugge:2017kzv}
J.~Feldbrugge, J.-L. Lehners, and N.~Turok, ``{Lorentzian Quantum Cosmology},''
  \href{http://dx.doi.org/10.1103/PhysRevD.95.103508}{{\em Phys. Rev.} {\bf
  D95} (2017) no.~10, 103508},
\href{http://arxiv.org/abs/1703.02076}{{\tt arXiv:1703.02076 [hep-th]}}.
%%CITATION = ARXIV:1703.02076;%%.

\bibitem{Feldbrugge:2017fcc}
J.~Feldbrugge, J.-L. Lehners, and N.~Turok, ``{No smooth beginning for
  spacetime},'' \href{http://dx.doi.org/10.1103/PhysRevLett.119.171301}{{\em
  Phys. Rev. Lett.} {\bf 119} (2017) no.~17, 171301},
\href{http://arxiv.org/abs/1705.00192}{{\tt arXiv:1705.00192 [hep-th]}}.
%%CITATION = ARXIV:1705.00192;%%.

\bibitem{DiazDorronsoro:2017hti}
J.~Diaz~Dorronsoro, J.~J. Halliwell, J.~B. Hartle, T.~Hertog, and O.~Janssen,
  ``{Real no-boundary wave function in Lorentzian quantum cosmology},''
  \href{http://dx.doi.org/10.1103/PhysRevD.96.043505}{{\em Phys. Rev.} {\bf
  D96} (2017) no.~4, 043505},
\href{http://arxiv.org/abs/1705.05340}{{\tt arXiv:1705.05340 [gr-qc]}}.
%%CITATION = ARXIV:1705.05340;%%.

\bibitem{Feldbrugge:2017mbc}
J.~Feldbrugge, J.-L. Lehners, and N.~Turok, ``{No rescue for the no boundary
  proposal: Pointers to the future of quantum cosmology},''
  \href{http://dx.doi.org/10.1103/PhysRevD.97.023509}{{\em Phys. Rev.} {\bf
  D97} (2018) no.~2, 023509},
\href{http://arxiv.org/abs/1708.05104}{{\tt arXiv:1708.05104 [hep-th]}}.
%%CITATION = ARXIV:1708.05104;%%.

\bibitem{Vilenkin:1982de}
A.~Vilenkin, ``{Creation of Universes from Nothing},''
\href{http://dx.doi.org/10.1016/0370-2693(82)90866-8}{{\em Phys. Lett.} {\bf
  117B} (1982)  25--28}.
%%CITATION = PHLTA,117B,25;%%.

\bibitem{Vilenkin:1984wp}
A.~Vilenkin, ``{Quantum Creation of Universes},''
\href{http://dx.doi.org/10.1103/PhysRevD.30.509}{{\em Phys. Rev.} {\bf D30}
  (1984)  509--511}.
%%CITATION = PHRVA,D30,509;%%.

\bibitem{Vilenkin:1994rn}
A.~Vilenkin, ``{Approaches to quantum cosmology},''
  \href{http://dx.doi.org/10.1103/PhysRevD.50.2581}{{\em Phys. Rev.} {\bf D50}
  (1994)  2581--2594},
\href{http://arxiv.org/abs/gr-qc/9403010}{{\tt arXiv:gr-qc/9403010 [gr-qc]}}.
%%CITATION = GR-QC/9403010;%%.

\bibitem{Vilenkin:2018dch}
A.~Vilenkin and M.~Yamada, ``{Tunneling wave function of the universe},''
  \href{http://dx.doi.org/10.1103/PhysRevD.98.066003}{{\em Phys. Rev.} {\bf
  D98} (2018) no.~6, 066003},
\href{http://arxiv.org/abs/1808.02032}{{\tt arXiv:1808.02032 [gr-qc]}}.
%%CITATION = ARXIV:1808.02032;%%.

\bibitem{Misner:1969hg}
C.~W. Misner, ``{Mixmaster universe},''
\href{http://dx.doi.org/10.1103/PhysRevLett.22.1071}{{\em Phys. Rev. Lett.}
  {\bf 22} (1969)  1071--1074}.
%%CITATION = PRLTA,22,1071;%%.

\bibitem{Halliwell:1988ik}
J.~J. Halliwell and J.~Louko, ``{Steepest Descent Contours in the Path Integral
  Approach to Quantum Cosmology. 1. The De Sitter Minisuperspace Model},''
\href{http://dx.doi.org/10.1103/PhysRevD.39.2206}{{\em Phys. Rev.} {\bf D39}
  (1989)  2206}.
%%CITATION = PHRVA,D39,2206;%%.

\bibitem{Gielen:2015uaa}
S.~Gielen and N.~Turok, ``{Perfect Quantum Cosmological Bounce},''
  \href{http://dx.doi.org/10.1103/PhysRevLett.117.021301}{{\em Phys. Rev.
  Lett.} {\bf 117} (2016) no.~2, 021301},
\href{http://arxiv.org/abs/1510.00699}{{\tt arXiv:1510.00699 [hep-th]}}.
%%CITATION = ARXIV:1510.00699;%%.

\bibitem{Gielen:2016fdb}
S.~Gielen and N.~Turok, ``{Quantum propagation across cosmological
  singularities},'' \href{http://dx.doi.org/10.1103/PhysRevD.95.103510}{{\em
  Phys. Rev.} {\bf D95} (2017) no.~10, 103510},
\href{http://arxiv.org/abs/1612.02792}{{\tt arXiv:1612.02792 [gr-qc]}}.
%%CITATION = ARXIV:1612.02792;%%.

\bibitem{DiTucci:2018fdg}
A.~Di~Tucci and J.-L. Lehners, ``{Unstable no-boundary fluctuations from sums
  over regular metrics},''
\href{http://arxiv.org/abs/1806.07134}{{\tt arXiv:1806.07134 [gr-qc]}}.
%%CITATION = ARXIV:1806.07134;%%.

\end{thebibliography}\endgroup

\end{document}